\documentstyle[12pt]{article}
\input{epsf}

\def\ie{\hbox{\it i.e.}}	
\def\mdot{\hskip -.1cm \cdot \hskip -.1cm}
\def\eg{\hbox{\it e.g.}}	
\def\etal{\hbox{\it et al.}}
\def\abs#1{\left| #1\right|}
\def\ltap{\raisebox{-.4ex}{\rlap{$\sim$}} \raisebox{.4ex}{$<$}}

\def\ap#1#2#3{           { Ann. Phys. (NY) }{\bf #1}, #2 (19#3)}
\def\ar#1#2#3{     { Ann. Rev. Nucl. and Part. Sci. }{\bf #1}, #2 (19#3)}
\def\np#1#2#3{           { Nucl. Phys. }{\bf #1}, #2 (19#3)}
\def\pl#1#2#3{           { Phys. Lett. }{\bf #1}, #2 (19#3)}
\def\pr#1#2#3{           { Phys. Rev. }{\bf #1}, #2 (19#3)}
\def\prl#1#2#3{          { Phys. Rev. Lett. }{\bf #1}, #2 (19#3)}

\begin{document}
\begin{titlepage}
\today          \hfill
\begin{center}
\hfill    LBL-37014 \\

\vskip .5in

{\large \bf Summary talk: Gauge Boson Self Interactions}
\footnote{This work was supported by the Director, Office of Energy
Research, Office of High Energy and Nuclear Physics, Division of High
Energy Physics of the U.S. Department of Energy under Contract
DE-AC03-76SF00098.}\footnote{Invited talk given at the International
Symposium on Vector Boson Self Interactions, UCLA, February 1-3 1995}

\vskip .50in

\vskip .5in
Ian Hinchliffe\\

{\em Theoretical Physics Group\\
    Lawrence Berkeley Laboratory\\
      University of California\\
    Berkeley, California 94720}
\end{center}

\vskip .5in

\begin{abstract}
A review is given of the theoretical expectations of the self couplings of
gauge bosons and of the present experimental information on the couplings.
 The possibilities for
future measurements are also discussed.

\end{abstract}
\end{titlepage}
%THIS PAGE (PAGE ii) CONTAINS THE LBL DISCLAIMER
%TEXT SHOULD BEGIN ON NEXT PAGE (PAGE 1)
\renewcommand{\thepage}{\roman{page}}
\setcounter{page}{2}
\mbox{ }

\vskip 1in

\begin{center}
{\bf Disclaimer}
\end{center}

\vskip .2in

\begin{scriptsize}
\begin{quotation}
This document was prepared as an account of work sponsored by the United
States Government. While this document is believed to contain correct
 information, neither the United States Government nor any agency
thereof, nor The Regents of the University of California, nor any of their
employees, makes any warranty, express or implied, or assumes any legal
liability or responsibility for the accuracy, completeness, or usefulness
of any information, apparatus, product, or process disclosed, or represents
that its use would not infringe privately owned rights.  Reference herein
to any specific commercial products process, or service by its trade name,
trademark, manufacturer, or otherwise, does not necessarily constitute or
imply its endorsement, recommendation, or favoring by the United States
Government or any agency thereof, or The Regents of the University of
California.  The views and opinions of authors expressed herein do not
necessarily state or reflect those of the United States Government or any
agency thereof, or The Regents of the University of California.
\end{quotation}
\end{scriptsize}

\vskip 2in

\begin{center}
\begin{small}
{\it Lawrence Berkeley Laboratory is an equal opportunity employer.}
\end{small}
\end{center}

\newpage
\renewcommand{\thepage}{\arabic{page}}
\setcounter{page}{1}

The electro-weak gauge bosons in the standard model of electroweak interactions
interact with
each other in a way that is fully described by the model. Deviations from the
prescribed form cause the model to be non-renormalizable or, equivalently, to
violate unitarity in high energy scattering \cite{smith}.
In this review talk, I shall present a personal perspective on the
determination
of, and expectations for, these couplings. I shall discuss the form of the
deviations from the standard model
 and how they are parameterized and then discuss the expectations for
the deviations in extensions to the standard model. I will review the current
experimental information and the possible impact of future
experiments.

Deviations from the standard model must be
parameterized in some way that will still allow predictions for experimental
quantities to be made.
It is convenient to begin with the general form of the $WWV$ coupling
 where $V$  is either a $Z$ boson or a photon \cite{peccei}.
\begin{eqnarray}
L/(ig_v)&=&(W_{\mu\nu}^{a\dagger}W^{a\mu}-W_{\mu}
^{a\dagger}W^{a\mu}_\nu)V^\nu g_1^V+
\kappa_VW_\mu^{a\dagger}W^a_\nu V^{\mu\nu}\nonumber \\
&+&\widetilde{\kappa_V}W_\mu^{a\dagger}W^a_\nu V_{\alpha\beta
}\epsilon^{\mu\nu\alpha\beta}
+\frac{\lambda_V}{M_W^2}W_{\rho\mu}^{a\dagger}W^{a\mu_\nu} V^{\rho\nu}
-ig_5^V \epsilon^{\mu\nu\rho\sigma}(W_\mu^{a\dagger}
\stackrel{\leftrightarrow}{\partial_\rho} W_\nu^{a\dagger})V_\sigma\nonumber \\
&+&\frac{\widetilde{\lambda_V}}{M_W^2}W_{\rho\mu}^{a\dagger}W^{a\mu}_\nu
 V_{\alpha\beta}\epsilon^{\rho\nu\alpha\beta}
+ig_4^V W_\mu^{a\dagger} W^a_\nu (\partial^{\mu}V^{\nu}+\partial^{\nu}
V^{\mu})
\label{wgamma}
\end{eqnarray}
$W^a_\mu$ ($W^a_{\mu\nu} $) represents the $W$ boson field (field strength) and
$V_\mu$ (or $V_{\mu\nu}$) is that of the photon ($\gamma$) or $Z$ boson.
The $SU(2)$ index $a$ will be dropped in what follows.
Electromagnetic gauge invariance implies that $g_5^\gamma=g_4^\gamma=0$.
In the standard model, $\lambda_Z=\lambda_\gamma=g_5^\gamma=g_5^Z=g_4^A=g_4^Z=
\widetilde{\kappa_V}=\widetilde{\lambda_V}=0$,
$\kappa_Z=\kappa_\gamma=g_1^Z=g_1^\gamma=1$,
$g_Z=e\cot\theta_W$ and $g_\gamma=e$.
Radiative corrections can induce small changes in these values at higher order
in perturbation theory.
The terms $\widetilde{\kappa}, \widetilde{\lambda}$ and $g_4$
violate CP and are also zero at
one loop in the standard model.
Experimental constraints are often quoted in terms of $\lambda$
and $\Delta\kappa=\kappa-1$ which parameterize deviations from the standard
model.  The other possible self couplings are $ZZZ$,
$ZZ\gamma$ and $Z\gamma\gamma$.
In the standard model these are zero. They are severely constrained by
electromagnetic gauge invariance and Bose symmetry and must vanish if all
of the particles are on mass shell \cite{peccei,berger}.
I will phrase most of the following discussion in terms of $\kappa_\gamma$ and
$\lambda_\gamma$ assuming that all the other couplings have the form given by
the standard model. The arguments provided below can be extended to the other
cases straightforwardly.

The standard, $SU(2)\times U(1)$ model, of electro-weak corrections has now
been
tested at the quantum (1-loop) level in experiments at LEP, SLC and
elsewhere\cite{dorothea,takeuchi}.
In these radiative corrections, the gauge
boson self interactions can appear in loop corrections to the $W$, $Z$ and
photon propagators. If all loops involving gauge boson self interactions  are
ignored, the agreement between
theory and experiment is less good \cite{sirlin,hagiwara}.
Direct determination of these self interactions comes from direct observation
of
gauge boson pairs at the Tevatron or, eventually, at LEPII.

Extensions to the standard model can produce values of the parameters in
Equation \ref{wgamma} that deviate from the standard model form. I will assume
that whatever extensions exist, they must satisfy $SU(2)\times U(1)$ gauge
invariance. A model that does not do this will be difficult to reconcile with
current data\footnote{For more discussion of  this see the talk by Willenbrock
at this meeting \cite{willen}}.
It is convenient to distinguish two types of extensions to the standard model.
First, there are models that, like the standard model, are renormalizable. In
this case a finite number of new parameters is sufficient to fully describe the
theory. Supersymmetric extensions of the standard model usually fall into this
class. In models of this type the parameters in Equation \ref{wgamma} are
modified by radiative (loop) corrections from the standard model values.

Second there are non-renormalizable theories. Such models have a mass
scale $\Lambda$ that appears in the coefficient of the higher dimension
operators. For experiments that probe energy scales ($E$)
 less than $\Lambda$, the
effects of these operators are suppressed by powers of $(E/\Lambda)$. Although,
such models contain, in principle, an infinite number of parameters, only a
few of these will be relevant for experiment since the suppression will render
the effects of most of them unobservable.
The theory can then be regarded as an effective
theory valid for $E<\Lambda$. At energies above $\Lambda$, the theory is
replaced by a more fundamental one and the terms in the effective theory are
computable in terms of the parameters of the more fundamental theory.
This notion of an effective theory is a
very useful one since it may be possible to severely constrain its form without
knowing the full dynamics of the fundamental theory \cite{georgi}. The best
example of this
type is the theory that describes the interaction of pions with each other at
low energy. Introducing $U=exp(i\overline{\sigma}
\mdot \overline{\pi}/f_\pi)$, where the vector $\overline{\pi}$
represents the $\pi^{\pm},\pi^0$, the interactions are given by
\begin{equation}
tr(\partial_\mu U^{\dagger}\partial_\mu U) + O(\frac{1}{4\pi f_\pi})^2
\end{equation}
This Lagrangian well describes QCD, \ie\ the dynamics of $\pi-\pi$ scattering,
on energy scales less than a few hundred
MeV.
At higher energies the full dynamics of (non-perturbative) QCD, including the
details or resonances is needed to fully describe the scattering.
The low energy Lagrangian is determined by the symmetries of low energy QCD,
\ie\ the fact that the pions are the Goldstone bosons of spontaneously broken
chiral symmetry.

If there is new dynamics on a mass scale of a few TeV, such as is the case in
technicolor\cite{techni}
models or models where there are strong interactions between
longitudinally polarized $W$ and $Z$ bosons at high energy\cite{strongw},
the effects of
this dynamics can be parameterized by adding terms to the standard model
Lagrangian\cite{wudka}.
 These form of these terms is dictated by the requirement that they
must not produce any effects that would invalidate the various standard model
tests and they must be invariant under $SU(2)\times U(1)$. The form of the
operators depends upon the particle content of the low energy effective theory.
The theory must contain the quarks, leptons and gauge bosons; it may or may not
contain Higgs scalars. If we assume that there are no light Higgs scalars then
one can write 12 CP invariant operators of dimension 4 \cite{longhitano} or
less. This lagrangian can be written as a gauged chiral model. In addition to
the quark and lepton fields and the gauge boson fields, there is a field
$\Sigma =exp(i\pi^a\tau^a/v)$ with $v=246$ GeV. The field $\pi^a$ provides
the longitudinal degrees of freedom for the massive $W$ and $Z$ bosons.
The kinetic energy for the gauge bosons is given by
\begin{equation}  \frac{v^2}{4}tr(D_\mu\Sigma^{\dagger}D^\mu\Sigma)-\frac{1}{2}
W_{\mu\nu}W^{\mu\nu}-\frac{1}{2}B_{\mu\nu}B^{\mu\nu}
\end{equation}
Here field $B_{\nu\mu}$  ($W_{\mu\nu}$)
is the field strength of the  $U(1)$ ($SU(2)$) part of the standard model.
These terms also give the mass for the $W$ and $Z$ bosons and the
photon. I will consider the effects of two of the additional operators
\begin{eqnarray}
L_1&=-&\frac{v^2}
{\Lambda^2}2ig\beta_1 tr(W_{\mu\nu}D^\mu\Sigma^{\dagger}D^\nu\Sigma)\nonumber\\
L_2&=&\frac{v^2}{\Lambda^2}g^2\tan\theta_W\beta_2(\Sigma
B_{\mu\nu}\Sigma^{\dagger}W^{\mu\nu})
\label{strong}
\end{eqnarray}
These give a contribution to $\kappa_\gamma$
\begin{eqnarray}
\Delta\kappa_\gamma&=&\frac{v^2}{\Lambda^2}g^2(\beta_1-\beta_2)\nonumber\\
\lambda_\gamma&=&0
\end{eqnarray}
However the term $L_2$ also contributes to the two point function of the gauge
bosons and is therefore constrained by measurements at LEP and elsewhere as I
will now discuss.

Recall how tests of the standard model are carried out.
The model is fully described in terms of a set of parameters which can be
taken, to be the Fermi constant $G_F$, the fine structure constant
$\alpha_{em}$, the
mass of the $Z$, the Higgs mass and the masses of the quarks and leptons.
Taking these values as input, one computes the expected value of some
experimentally
observable quantity such as the cross section of $\nu-e$ scattering. This
expected quantity has some error $\delta_{theory}$, that arises from the
uncertainties in the parameters and residual uncertainty arising from the
the calculation having been carried
out to some order in perturbation theory. This is then compared with an
experimental measurement which has an error $\delta_{expt}$. If the theory and
experiment agree, the model is the tested with an accuracy that is the
{\it larger} of $\delta_{theory}$ and $\delta_{expt}$.
 A failure of the model is
revealed when there are experimental results that disagree with theory by more
than the
{\it larger} of $\delta_{theory}$ and $\delta_{expt}$.
In a variant of the standard
model, extra parameters appear and the values of these parameters can be
adjusted to accommodate experimental values that the standard model fails to
predict correctly.

The parameters appearing in equation \ref{wgamma} need to be related to
physical quantities so that their values can be extracted from data. The
general form of the $WWV$ vertex for bosons of momenta $p_1$, $p_2$ and $p_3$
and polarization tensors
$\epsilon_1^\mu$, $\epsilon_2^\nu$ and $\epsilon_3^\alpha $
depends upon the invariant mass of the three
bosons {\it viz.} $\Gamma^{\mu\nu\alpha}(p_1^2,p_2^2,p_3^2)$. In the case of
the $WW\gamma$ vertex, there is a physical point where all of the particles are
on mass shell (static limit) \ie \ $\Gamma^{\mu\nu\alpha}(M_W^2,M_W^2,0)$. At
 this point the
quantities appearing in equation \ref{wgamma} are related to physical
properties of the $W$
boson; $\kappa_\gamma$ and $\lambda_\gamma$ to the
electric quadrapole moment ($Q$)  and magnetic dipole moment ($\mu$) of the
$W$.
\begin{eqnarray}
\mu&=&\frac{e}{2M_W}(1+\kappa_\gamma+\lambda_\gamma)\\
Q&=&-\frac{e}{M_W^2}(\kappa_\gamma-\lambda_\gamma)\\
\end{eqnarray}
However these static quantities are not sufficient
to describe the general properties of
$\Gamma^{\mu\nu\alpha}(p_1^2,p_2^2,p_3^2)$.

Consider the process $q\overline{q}\to W\gamma$; I will assume for simplicity
that all of the parameters in the $WW\gamma$ vertex have the standard form
except for $\kappa_\gamma$ and $\lambda_\gamma$.
\begin{figure}
\epsfbox{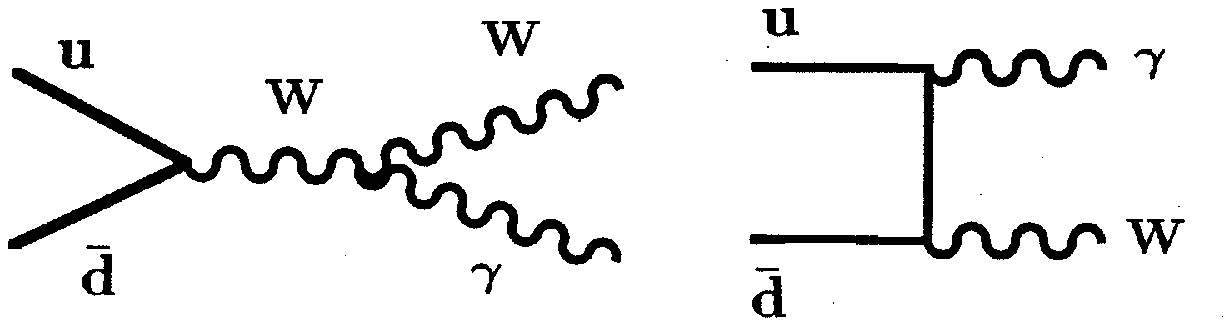}
\label{qqtowgam}
\caption[]{Feynman diagrams showing the process $q\overline{q}\to W\gamma$}
\end{figure} There is a contribution for
the Feynman diagram shown in figure \ref{qqtowgam} which depends on
$\Gamma^{\mu\nu\alpha}(s,M_W^2,0)$ where $\sqrt{s}$
is the center of mass energy of
the quark antiquark system. If $\kappa_\gamma$ and $\lambda_\gamma$ are taken
to be
constants, then this will result in a scattering amplitude of the form
\begin{equation}
A\sim a+b \sqrt{s} (\kappa_\gamma-1+\lambda_\gamma) +c s\lambda_\gamma
\label{unitarity}
\end{equation}
where $a$, $b$ and $c$ are independent of the center of mass energy
 ($\sqrt{s}$).
This amplitude grows with $s$ unless $\kappa_\gamma$ and $\lambda_\gamma$
 have the standard
model values of $1$ and $0$ respectively. This growth is a general feature of
anomalous couplings. It is immediately clear that the sensitivity of  an
experiment to the anomalous couplings increases with the energy of the
experiment and that a
high energy experiment is more sensitive to $\lambda_\gamma$ than to
$\kappa_\gamma-1$. Hence an $e^+e^-\to W^+W^-$
measurement at $\sqrt{s}\sim 500 $ GeV can constrain $\lambda_\gamma$
 and $\kappa_\gamma-1$
much more precisely than a measurement with comparable statistical power at
$\sqrt{s}\sim 190$ GeV. Similarly in a hadron collider, the greatest
sensitivity
arises from the (few) events of largest energy.

This problem of unitarity violations can be avoided
phenomenologically by the introduction of form factors \cite{zeppen}
 to damp the growth at
large $s$ \ie \  $\lambda_\gamma \to \lambda_\gamma/(1+s/\Lambda^2)^{n_1}$
and $(\kappa_\gamma-1)
 \to (\kappa_\gamma-1)/(1+s/\Lambda^2)^{n_2}$ with $n_1,n_2\ge 1$.
It is conventional to use a dipole form factor, \ie\ $n_2=n_1=2$.
 An experiment measuring
the $W\gamma$ production cross section can
set a limit on $\Delta\kappa_\gamma$ and $\lambda_\gamma$
 given a value of $n_1$, $n_2$, and
$\Lambda$. Note that for a given choice of $n_1$, $n_2$, and
$\Lambda$, unitarity alone bounds $\lambda_\gamma$ and
 $\kappa_\gamma-1$.
For $n_1=n_2=2$, this bound is \cite{zeppen1}
\begin{eqnarray}
\abs{\kappa_\gamma-1}&<&7.4 \hbox{ TeV}^2/\Lambda^2
\nonumber\\
\abs{\lambda_\gamma}&<&4.0 \hbox{ TeV}^2/\Lambda^2
\end{eqnarray}
An experiment that is not sensitive to values below these
is not relevant.

\begin{figure}
\epsfbox{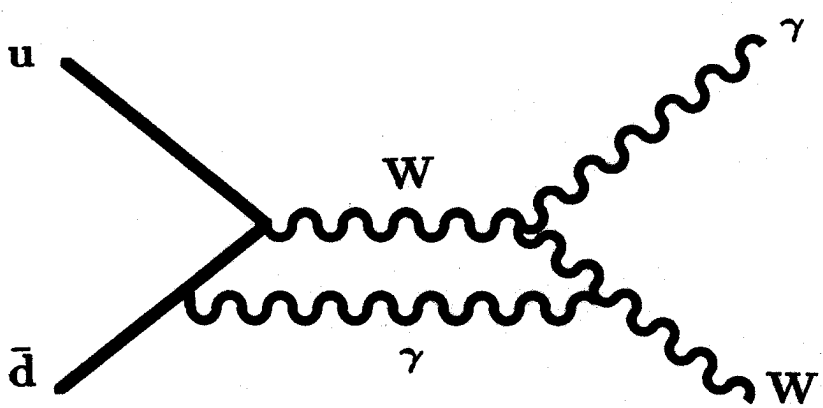}
\caption[]{An example of contribution to $q\overline{q}\to W\gamma$ which must
be
included along with the 1 loop corrections to the $WW\gamma$ vertex appearing
in Figure \ref{qqtowgam}}
\label{figprop}
\end{figure}

Generally $\Gamma^{\mu\nu\alpha}(p_1^2,p_2^2,p_3^2)$ is not gauge invariant
when computed beyond leading order in perturbation theory. This is
directly related to the fact that it is not a physical quantity.  As discussed
in reference \cite{pinch},
it is possible to define a gauge invariant form by including
some pieces of other  corrections that would contribute at the same order
in perturbation theory to a physical process. In the example of
$q\overline{q}\to W\gamma$, a contribution of this type is shown in figure
\ref{figprop}.
It is convenient to quote the values of the physical quantities
$\lambda_\gamma$
and $\Delta\kappa_\gamma$ at the static limit
as a measure of the expected size of the higher order corrections.

What values of anomalous couplings are to be expected in the standard model and
its possible extensions?  In the standard model the natural
size of $\kappa_\gamma-1$ and $\lambda_\gamma$ is $\alpha_{em}/\pi$
\cite{bardeen}. For a top
quark mass of 150 GeV and a Higgs mass of 100 GeV, $\lambda_\gamma=0.006$ and
$\kappa_\gamma+\lambda_\gamma-1=-0.0003$\cite{agres}.
In the supersymmetric
model the size of the corrections depends upon the masses of the supersymmetric
particles. Note that the masses assumed must be consistent with other
experimental constraints. For most of the values of the parameters,
$\lambda_\gamma$ is
about 60\% of its value in the standard model and
$\kappa_\gamma+\lambda_\gamma-1$ is about 5
times larger than its standard model value.

In extensions to the standard model where operators of the type in equation
\ref{strong} are present, we need to estimate the size of  $\beta_1, \beta_2$,
and $\Lambda$. Using the scale of new physics $\Lambda$ to
be 1 TeV we might expect  $\delta\kappa_\gamma $ to be as large as 0.05 if
$\beta_1\sim\beta_2\sim 1$ as would be expected if the new physics at scale
$\Lambda $ is strongly coupled. Other estimates yield values smaller than these
\cite{wudka}. The term $L_2$ in equation \ref{strong} contributes to the
gauge boson two point functions and in particular to the Peskin-Takeuchi
\cite{peskin} $S$ parameter. Using the data from LEP, the constraint
$\abs{\beta_2}\ltap 0.5$\cite{dawson}  is obtained (again I have taken
$\Lambda=1$ TeV). Hence the contribution of $L_2$, to $\Delta\kappa_\gamma$
is restricted to be less than 0.013. The term $L_1$ is not directly constrained
by LEP data. However since both $\beta_1$ and $\beta_2$ arise from the same
(unknown) physics, it is to be expected that they will be of the same order of
magnitude.

\begin{figure}
\epsfbox{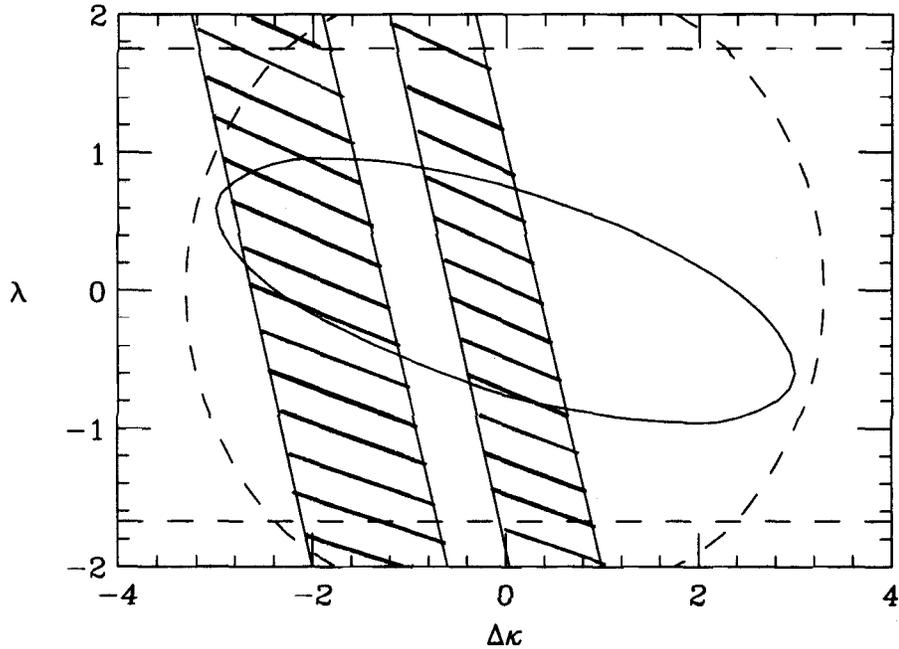}
\caption[]{The limits on $\Delta\kappa_\gamma$ and $\lambda_\gamma$ from the D0
experiment (the area inside the oval region is allowed region) \cite{d0}.
 The limits from CDF are similar \cite{cdf}.
Also shown is the allowed region from the observation of $b\to s\gamma$ (the
hatched area) from
CLEO\cite{cleob}. The limits are shown at 95\% confidence.
The area outside the dashed circle is excluded by unitarity for the process
$q\overline{q}\to W^+W^-$ with $n_1=n_2=2$ and $\Lambda=1.5$ TeV. The regions
at the top and bottom of the figure bounded by the dashed horizontal
lines are excluded by unitarity in $q\overline{q}\to W_\gamma$}
\label{limits}
\end{figure}

There have been observations of $W\gamma$,
$Z\gamma$, $WW$ and $WZ$ final states at the
Tevatron collider by both CDF\cite{cdf} and D0\cite{d0} that are reviewed at
this meeting \cite{aihara,fuess}
The former constrains the $WW\gamma$
vertex while the latter constrains the $ZZ\gamma$ and $Z\gamma\gamma$ vertices
and the last constrain $WWZ$ and $WW\gamma$ vertices.
The limits on $\kappa_\gamma$ and $\lambda_\gamma$ arising from observation of
$W\gamma$ final states are shown in Figure
\ref{limits}. These limits use dipole form factors ($n_1=n_2=2$) with
$\Lambda=1.5$ TeV. The limits are essentially unchanged if $\Lambda=1$ TeV.
The unitarity limits for $\Lambda=1.5$ TeV are larger than the experimental
constraints (see figure \ref{limits}).

The limits on $\kappa_Z$
and $\lambda_Z$ arising from the observation of $WW$ and $WZ$ final states is
similar to those on $\kappa_\gamma$ and $\lambda_\gamma$ \cite{cdf}.
In the case of the $ZZ\gamma$ and $Z\gamma\gamma$ vertices, the limits are more
sensitive to the assumed form factor behaviour of the vertices \cite{berger}.
This is due to the form of the vertex function,
$\Gamma^{\mu\nu\alpha}(p_1^2,p_2^2,p_3^2)$, which must
vanish when the particles are all on mass shell and therefore  has powers of
 energy in the numerator. The form factors then introduced to prevent a
 unitarity violation must have $n\ge 3$.
Constraints have also been placed on the $ZZ\gamma$ couplings by searching for
events at LEP of the form $Z\to \gamma Z^*(\to \nu\overline{\nu})$
       \cite{l3}. These limits are comparable to those from CDF.

Note that the limits depend upon the ability to predict the event rates given
the gauge boson self couplings requires an understanding of the QCD production
process. This process is computed at next to leading order in $\alpha_{strong}$
and the resulting uncertainty should quite small \cite{ohnemus}. The angular
distribution of the process $q\overline{q}\to W\gamma$ has a zero at a
particular value of the scattering angle \cite{brown}. This zero is not
preserved by the higher order QCD corrections.

\begin{figure}
\epsfbox{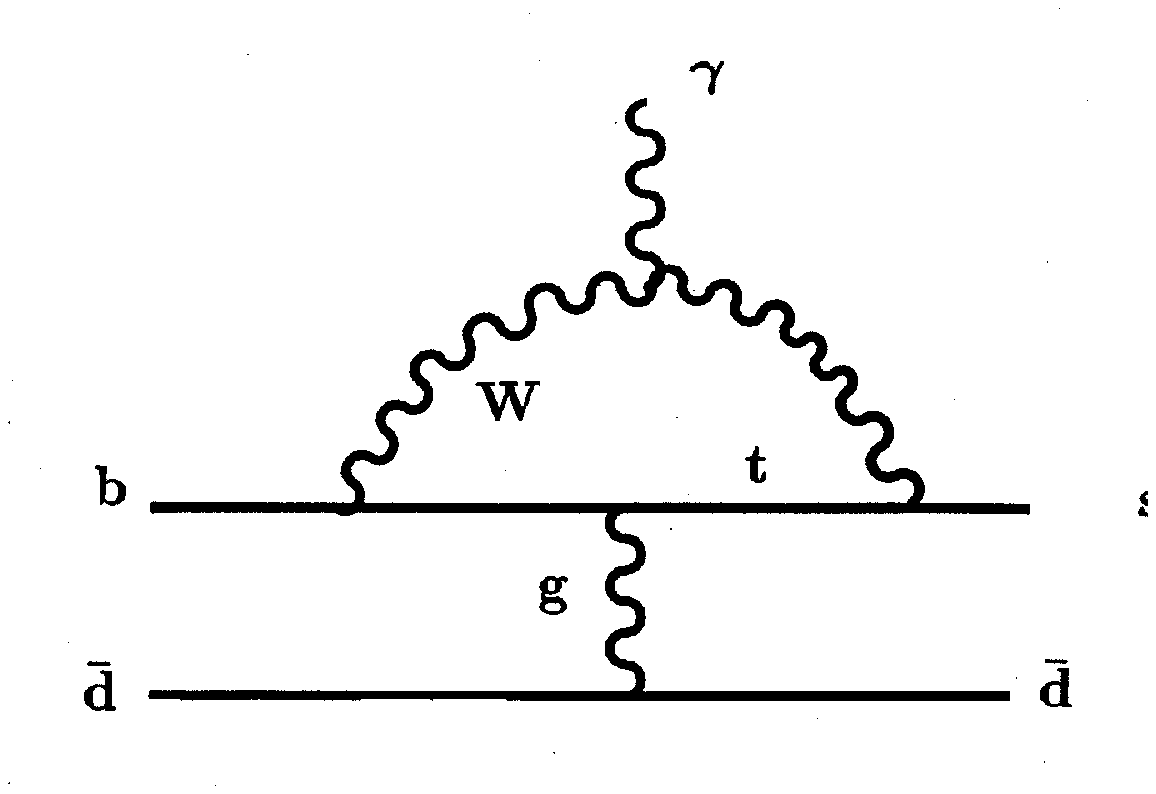}
\caption[]{A contribution to the process $B\to K\gamma$}
\label{btowgam}\end{figure}

The decay of a B meson to a photon and a strange meson, proceeds via loop
effects. One relevant graph is given in Figure \ref{btowgam}, where the
$WW\gamma$
vertex is present.
The experimental observation of this process \cite{cleob}
enables one to constrain $\kappa_\gamma$ and $\lambda_\gamma$
\cite{bsgamtheory}.
The constraint is shown on Figure \ref{limits}. Note that the constraint is
less direct than that of CDF and D0. The interference between the graph shown
in figure \ref{btowgam} and other graphs such as the one where the photon is
radiated off the top quark, results in the odd shape for the allowed region.
If there were other diagrams that could
contribute to $b\to s\gamma$,  such as would occur in a supersymmetric model,
the constraint becomes a coupled limit involving the couplings of other
particles \cite{hewett}.

I will end with a discussion of the prospects for future measurements.
LEP II will be able to measure the  $Z\gamma$ and $WW$ and possibly the $ZZ$
final state. Consequently it will probe the $WW\gamma$, $ZZ\gamma$,
$Z\gamma\gamma$ and $WWZ$ vertices. In the case of $WW\gamma$, the sensitivity
of order 0.3 (0.5) to both $\lambda$ and $\Delta\kappa$ at $\sqrt{s}=192
(176)$ GeV \cite{lepii}.
 This is approximately three times better than the current limits
from the Tevatron. However these limits are based on $\sim 15$ pb$^{-1}$
of data. They will improve by the end of the current when $\sim 100$
pb$^{-1}$ will be available. If it is then possible to combine the CDF and D0
limits, they should fall by a factor of three or so.
 It seems reasonable to conclude therefore that any improvement that LEP II
can provide over the Tevatron will be small.

There has been much discussion in the literature \cite{dawson,hernandez}
and at this meeting of the
extent to which the precision measurements of LEP imply that LEPII cannot
see any effects of anomalous couplings. In order to address this question,
possible models that differ from the
standard model must be constructed so that they are consistent with LEP data
and predictions for anomalous couplings or measurements at LEPII made. As
discussed above, the LEP data constrain $\beta_2$ of Equation \ref{strong}
sufficiently that the contribution of $L_2$ to anomalous couplings is too small
to be seen at LEPII.  The ``natural'' values of $\beta_2$ and $\beta_1$ should
be roughly equal.
In this case it is unlikely that LEPII (or the Tevatron) will see a positive
effect. However, it might happen that $\beta_1>>\beta_2$. In QED, one can
estimate the natural size of a process by assuming that the coefficient of the
appropriate power of $\alpha_{em}/\pi$ is order one. Large coefficients such as
$\pi^2$ that appears in the radiative corrections to Coulomb scattering
\cite{scat} as well as ones that are less than one, such as the order
$\alpha\pi$
correction to $g-2$ of the electron do occur.

The sensitivities of experiments discussed above are very far from the
deviations from the standard model that can reasonably be expected.
\footnote{A
participant asked me if this meant that theory obviated experiment.}
Experiments at LHC \cite{womersly,atlas} have greater sensitivity because of
their
greater energy. ATLAS expects a sensitivity of order
 $\Delta\kappa_\gamma\sim 0.04$ $\lambda_\gamma\sim 0.0025$ which is
 approaching values that are theoretically interesting\cite{atlas}.
 An $e^+e^-$ collider
 with more energy than LEP will be more sensitive; at $\sqrt{s}=500$ GeV (1.5
 TeV) the sensitivities are $\lambda_\gamma$ and $\Delta\kappa_\gamma$ are
$\sim 0.01$ ($\sim 0.002$) \cite{barklow}.

I am grateful to the members of the organizing committee, U. Baur, S. Errede
and T. M\"{u}ller for their work in making this conference such a success.
The work was supported by the Director, Office of Energy Research,
Office of High Energy Physics, Division of High Energy Physics of the
U.S. Department of Energy under Contract DE--AC03--76SF00098.
Accordingly, the U.S. Government retains a nonexclusive, royalty-free
license to publish or reproduce the published form of this contribution,
or allow others to do so, for U.S. Government purposes.

\end{document}